# Ultralong Copper Phthalocyanine Nanowires with New Crystal Structure and Broad Optical Absorption

Hai Wang [1, ◊], Soumaya Mauthoor [2], Salahud Din [2], Jules A. Gardener [1, †], Rio Chang [2], Marc Warner [1], Gabriel Aeppli [1], David W. McComb [2], Mary P. Ryan [2], Wei Wu [1], Andrew J. Fisher [1], A. Marshall Stoneham [1], Sandrine Heutz [2, *]

[1] Department of Physics and Astronomy and London Centre for Nanotechnology, University College London, London WC1E 6BT, UK.
[2] Department of Materials and London Centre for Nanotechnology, Imperial College London, London SW7 2AZ, UK.
[◊] Present Address: Department of Physics, Kunming University, Kunming, 650214, P. R. China.
[†] Present Address: Department of Physics, Harvard University, Cambridge, MA 02138, USA.
[*] e-mail: s.heutz@imperial.ac.uk

The development of molecular nanostructures plays a major role in emerging organic electronic applications, as it leads to improved performance and is compatible with our increasing need for miniaturisation. In particular, nanowires have been obtained from solution or vapour phase and have displayed high conductivity[1, 2] or large interfacial areas in solar cells.[3] In all cases however, the crystal structure remains as in films or bulk, and the exploitation of wires requires extensive post-growth manipulation as their orientations are random. Here we report copper phthalocyanine (CuPc) nanowires with diameters of 10-100 nm, high directionality and unprecedented aspect ratios. We demonstrate that they adopt a new crystal phase, designated η-CuPc, where the molecules stack along the long axis. The resulting high electronic overlap along the centimetre length stacks achieved in our wires mediates antiferromagnetic couplings and broadens the optical absorption spectrum. The ability to fabricate ultralong, flexible metal phthalocyanine nanowires opens new possibilities for applications of these simple molecules.

Organic vapour phase deposition (OVPD) has been very successful at generating a wide range of film morphologies with a number of functional molecular materials, including smooth amorphous films,[4] textured islands,[5] and, more recently, nanowires.[6] The latter have been investigated for a variety of applications including sensors,[7] field effect transistors[1-2, 8] and photovoltaic cells.[3] So far, with the exception of some very short nanobrushes,[9] only wires randomly oriented on a substrate have been produced. Their implementation on a device therefore requires micromanipulation, which is severely limiting for future applications where texture and length are desirable.

Metal phthalocyanines (MPcs) are planar aromatic macrocyles which crystallise as a range of at least fifteen distinct polymorphs.[10, 11] Typically, thin films exist as either α or β-phases, which have also been observed in relatively thick ( > 100 nm in diameter) wires.[6] The optoelectronic[12] and magnetic[13] properties of Pc materials are strongly dependent on their structure, and the creation of a new polymorph can have important consequences both for devices and in industry where α, β and ε-CuPc forms are routinely used as pigments.

Through the optimisation of growth parameters in the OVPD we have obtained CuPc nanowires with diameters of 10-100 nm for lengths up to 1.4 cm. They adopt a new structure, with novel electronic absorption and magnetic properties. Their high directionality could significantly facilitate parallel integration into devices, and leads to confined antiferromagnetically coupled spin chains along the long wire axis.

Figure 1.a shows a photograph of the CuPc branches after a growth time of typically 120 minutes. Those branches nucleate in a region of the sample tube which is ~ 10 cm outside the furnace, where the temperature is close to ambient. The branches maintain their radial growth direction inwards until they reach the opposite side of the quartz tube, corresponding to maximum lengths of 1.4 cm. Similar growth patterns were obtained on polyimide (Kapton) substrates rolled inside the inner tube, yielding nanowire deposition over areas of about 60 $cm^2$. Deposition on non-flexible substrates such as glass or silicon is also possible, although there the nucleation occurs preferentially at the edge of the substrate facing the flow (Figure 1.b). Figures 1.c and d show that the branches contain bundles of nanowires maintaining the same long-range directionality as their parent branches. To our knowledge, this is the first report of CuPc branches made of nanowires that maintain a common growth direction to produce such a high aspect ratio. This is most probably due to their new structure as well as to the size and anisotropic adsorption and growth of the nuclei, as will be discussed below. There is no discernible variation in diameters of the nanowires along their length, implying one-dimensional growth from a nucleation site, without any significant build-up on the sides as deposition progresses.

The optical properties of CuPc solids are closely related to the relative orientation and separation of adjacent molecules[14] and are therefore specific to particular crystal structures. Indeed, the Q-band characteristic of single molecules in solution (upper panel in Figure 2a) is split by intermolecular interactions in the solid state and additional charge-transfer bands also contribute to the spectrum of thin films.[15] The spectrum of the wires displays two broad peaks at ~ 615 nm and ~ 765 nm, which is very different to what is observed for the known α and β-phases. The absorption of the wires cannot be simulated with any linear combination of α, β and monomer contributions, and therefore the wires must adopt a new structure. Models have been developed to correlate the intermolecular shifts and the splitting of the absorption peaks,[16] but those are not sufficient to unambiguously predict crystal structure based on absorption spectra. Therefore, we can only conclude that the increased splitting is derived from enhanced intermolecular hopping, which is advantageous for applications such as solar cells where harvesting light, especially in the infra-red, and then transferring charge is a challenging goal.

The overall crystallinity of the nanowires was determined using X-ray diffraction, Figure 2b, for a range of substrates as described in Methods. Identical peak positions are obtained in all cases, but intensity ratios vary, due to changes in the orientation of the wires as they are collected on the substrates. The main reflections are summarised in Table 1 and while the two lowest angle peaks are frequently observed in thin film data as the (100) and (001) reflections in the α-phase,[11] none of the structures reported in the literature can rationalise a lattice spacing of 10.2 Å (and its associated second order peak at 5.1 Å). We therefore again conclude that the nanowires are a new polymorph of CuPc.

Transmission electron microscopy (TEM), Figures 3a-c, shows that wire diameters can be as low as 10 nm. Average sizes are around 50 nm, in agreement with dimensions derived from XRD peak width using the Debye-Scherrer equation.[17] Observation of lattice fringes in the TEM images highlights the crystalline nature of the wires. The fringes can either be very straight or slightly wavy, and we attribute them to different groups of planes (001) and (200) respectively, as determined following indexation below. The waviness could originate either from out-of-plane buckling or the formation of multiple twin boundaries (as observed previously in STM[18]). The edges of some wires are curved with no fringes, possibly as a result of less-ordered growth at the grain edge. The intermittent loss of fringes within the wires in all figures is also most probably due to buckling, while more severe misorientation can be responsible for the absence of fringes in a minority of wires.

Figure 3e-f (white spots) shows transmission electron diffraction patterns (DP) of the wires. Two types of DPs were obtained; similar to Figure 3e where both high and low angle reflections are observed, or to Figure 3f, where the high angle reflections are either blurred or absent. The associated lattice spacings can also be distinguished as distributions centred either on $12.0 \pm 0.1$ Å and $3.77 \pm 0.02$ Å or $12.6 \pm 0.2$ Å respectively.

Table 1 summarises the experimental planes observed through the diffraction techniques described above. The systematic absences in the DPs are characteristic of the $P2_1/c$ space group of β-CuPc, which is used as a basis for indexation. Assigning the planes to the (001), (002), (20-1) and (010) reflections yields the unit cell parameters for the new polymorph, which we name η-CuPc, as summarised in the Table.

The molecular packing within the unit cells cannot be determined experimentally and was instead calculated by minimising the lattice potential of the proposed wire CuPc unit cell (see supplementary information). The structure converges to a molecular packing as described in Figure 3d (full atomic coordinates available in supplementary information). Simulations of the electron diffraction patterns along the respective axes calculated using SingleCrystal™ are shown in red (displaced for clarity) in Figures 3e-f. The agreement between experiment and simulations is very good, validating our model.

The mechanism for wire growth can be understood as a rapid nucleation due to supersaturation of the vapour after the sharp temperature drop at the exit of the furnace, followed by preferential deposition of the nuclei with the b-axis perpendicular to the chamber walls. Indeed, the (010) plane presents the highest aromatic contribution, which is different to the β-phase and brickstack α-phase. It is known that the stability of crystal phases is strongly correlated with their size,[19] which could be the driving force for forming nuclei of the η phase.

Figure 2d compares the magnetisation as a function of field at a temperature of 2K for the nanowires to that for the α and β-polymorphs.[13] The wires display weak antiferromagnetic coupling, intermediate between what we find for the paramagnetic β and antiferromagnetic α phases. The exchange interaction was derived from a fit of the field and temperature dependent magnetisation using the Bonner-Fisher model for Heisenberg spin-1/2 chain[20] and established to be $J_{ex}$ = - 0.5 K. We calculate the exchange interaction as described before[13] and find $J_{DFT}$ = - 1.8 K and $J_{PT}$= -0.4 (a.u., using identical proportionality constant as in ref. 13). DFT agrees with the sign of the interaction but the magnitude of the experimentally determined exchange splitting (~0.1 meV) challenges the accuracy of the method. The perturbation theory approach predicts trends correctly and rationalises the reduction in J by the smaller overlap between the lowest unoccupied molecular orbitals (LUMOs) of adjacent molecules in the wire phase relative to the α-phase (see schematics of the $e_g$-LUMO overlaps in Figure 4.d). Therefore, both our theoretical and experimental results highlight the specific magnetic functionality of the new wire phase which acts as a 1D quantum antiferromagnet.

While nanoscale features determine most physical properties, the exceptional length of the wires imparts them with large-scale flexibility, as demonstrated by a simple bending test (illustrated in Figure 2c). Characteristic spectroscopic properties are retained after mechanical stress, as seen in the lower panel of Figure 2a, highlighting the compatibility of the nanostructures with flexible plastic electronics.

Molecular nanowires with new structural, spectroscopic and magnetic properties have been produced by organic vapour phase deposition. The wire diameters can be as low as 10 nm across lengths exceeding 1.4 cm; the resulting aspect ratios (up to $10^6$) are to our knowledge the highest for a functional molecular crystal. Compared to carbon nanotubes, their closest morphological analogues, η-CuPc nanowires benefit from chemical purity (no catalyst) and additional functionality. The strong electronic overlap along the long axis of the wire mediates antiferromagnetic coupling and should also lead to high anisotropic conductivity, which could be exploited in field-effect and photovoltaic devices. The magnetic wire characteristics might also be valuable for quantum information processing and transmission,[21] which in strategies based on carbon nanotubes require doping. The unique directionality and flexibility of the wires combined with the new functionalities offers opportunities for direct integration into optimised nanoscale devices without the need for post-growth manipulation. Furthermore, fabrication of ferromagnetic (e.g. for MnPc or via doping) wires using the same self-assembly technique might enable devices based on shuttled domain walls such as racetrack memory[22] or classical magnetic logic.[23]

**Methods**

The OVPD system consists of a quartz tubular chamber (1.5 m long and 3 cm in diameter) inserted into a three-zone HZS-12/900E Carbolite furnace where the temperature of the individual zones (each 30 cm in length) can be independently controlled. The flow of the nitrogen carrier gas is controlled by a MKS 1179A Mass-Flo® Controller giving a rate of 350 sccm. The base pressure of the chamber is ~$3\times10^{-3}$ mbar before introducing nitrogen gas and increases to 6.0 mbar during the growth. A 1.2 m long quartz tube inside the chamber (inner diameter of 1.4 cm) acts as an anchoring point for the vertical growth of the nanowires and contains the source material, which is maintained in a boat at the centre of the hottest zone. The CuPc powder, (Aldrich 97%) was twice purified using gradient sublimation for the samples used in the magnetic measurements. The temperature of the furnace at the first, second and third zone during deposition is 480 °C, 440 °C and 180 °C respectively, and these values were reached through ramping from ambient at 20 °C/min until 40 °C below maximum temperature, followed by 10 minutes dwelling, slower ramp at 3 °C/min for 10 minutes, 5 minutes dwelling and final ramping at 1 °C/min. Growth times are reported from the moment the heating is started from room temperature conditions, so that the final source temperature is attained after a growth time of 60 minutes.

The CuPc nanowire samples were characterized by LEO-Gemini 1525 field emission gun scanning electron microscope (SEM), JEOL 2010 and JEOL 200FX transmission electron microscopes (TEM) operated at 200 kV, Philips PW1700 series diffractometer (using Cu Kα radiation) operated in the θ-2θ mode, Perkin Elmer Lambda UV/Vis spectrometer and a SQUID-based Magnetic Property Measurement System (Quantum Design). For TEM, the microgrid was moved until the beam was parallel with a low index crystal zone axis and a DP was observed. The possible misalignment might alter spot distances by up to 2%.

The nanowire samples were prepared for characterisation by attaching the CuPc nanowires to ethanol wetted substrates through capillarity. Substrates were carbon-coated copper grids for TEM or cleaned microscope slides for SEM and UV/Vis. The SEM samples were subsequently coated with a ~ 10 nm chromium layer to avoid electron charging and improve image quality. The nanowire samples for XRD were prepared by using a clean microscope glass slide or silicon substrate to collect CuPc branches from multiple experiments, thus a thick sample consisting of randomly oriented CuPc branches was measured. Samples were also prepared by growth of the wires directly on a silicon substrate, and their XRD scan is labelled "on substrate" in Figure 2b. Magnetic characterisation was performed in a similar manner to that published by our group,[13] i.e. by collecting the nanowires onto a thin strip of polyimide substrate (Kapton) and compensating with bare Kapton at either side of the film to subtract signal from the substrate.

**Acknowledgements**

We thank J. Hawley for preliminary experiments, A. Djurisic and A. Mostofi for useful discussions, and R. Sweeney and M. Ardakani for help with XRD and TEM respectively. Financial support from the Research Council UK and the Engineering and Physical Sciences Research Council (EPSRC) Basic Technology grants "Putting the Quantum into Information Technology" and "Molecular Spintronics" is gratefully acknowledged. We thank the Royal Society for a Dorothy Hodgkin Research Fellowship (SH) and a Royal Society Wolfson Research Merit Award (GA).

**Figure captions**

**Figure 1** **Photographic and SEM images of η-CuPc branches and nanowires.** Digital photographs showing the growth of CuPc branches inside the quartz tube (a) and on a glass substrate (b). SEM image of a single η-CuPc branch showing bundles of nanowires (c-d).

**Figure 2** **Spectroscopic, structural and magnetic characterisation of the nanowires** (a) Upper panel: Electronic absorption spectra of a range of CuPc samples, including solution, wires collected on a glass slide, α and β films. The wire absorption extends further in the red than any known thin film polymorphs. Lower panel: Electronic absorption of the wires on kapton both before and after the bending test illustrated in (c). (b) XRD scans of wires collected on glass and silicon after growth on the chamber walls, and of wires directly grown on a silicon substrate (labelled "on substrate"). Vertical marker lines are used to represent the (00n), (2n00) and (2n0-n) family of planes in dashed, dotted and continuous lines, while the sharper peak denoted by * is due to the silicon substrate. XRD scans of α and β powders highlight that the reflections at 8.8° cannot be seen in common polymorphs. (c) Bending test of the wires on kapton. (d) Magnetisation as a function of field at 2K for alpha, beta and wire samples. The curves were normalised at 7 Tesla; actual moments at those fields are close to 1 $\mu_B$ for β, 0.8 $\mu_B$ for α and intermediate between these values for η. The LUMOs involved in the indirect exchange are shown. In this convention, a positive contribution to the hopping matrix elements between orbitals on the two molecules is shown in grey, while a negative contribution is shaded in black.

**Figure 3** **TEM images of isolated single branches**. (a) Well-defined straight fringes due to the 001 planes, (b) more wavy fringes due to the 200 planes and (c) wires with a diameter of 10 nm. (d) Schematic of the unit cell of η-CuPc. Typical transmission electron diffraction patterns for wires for [100] and [001] projections (e and f respectively) and their corresponding simulated pattern offset for clarity. Experimental and simulated spots are shown in white and red respectively (displaced for clarity).

**Table 1** (a) Summary of experimental and calculated crystal lattice planes for wire structure, and (b) summary of lattice parameters for the wire structure, compared to $\alpha^{11}$ and $\beta^{24}$ values.

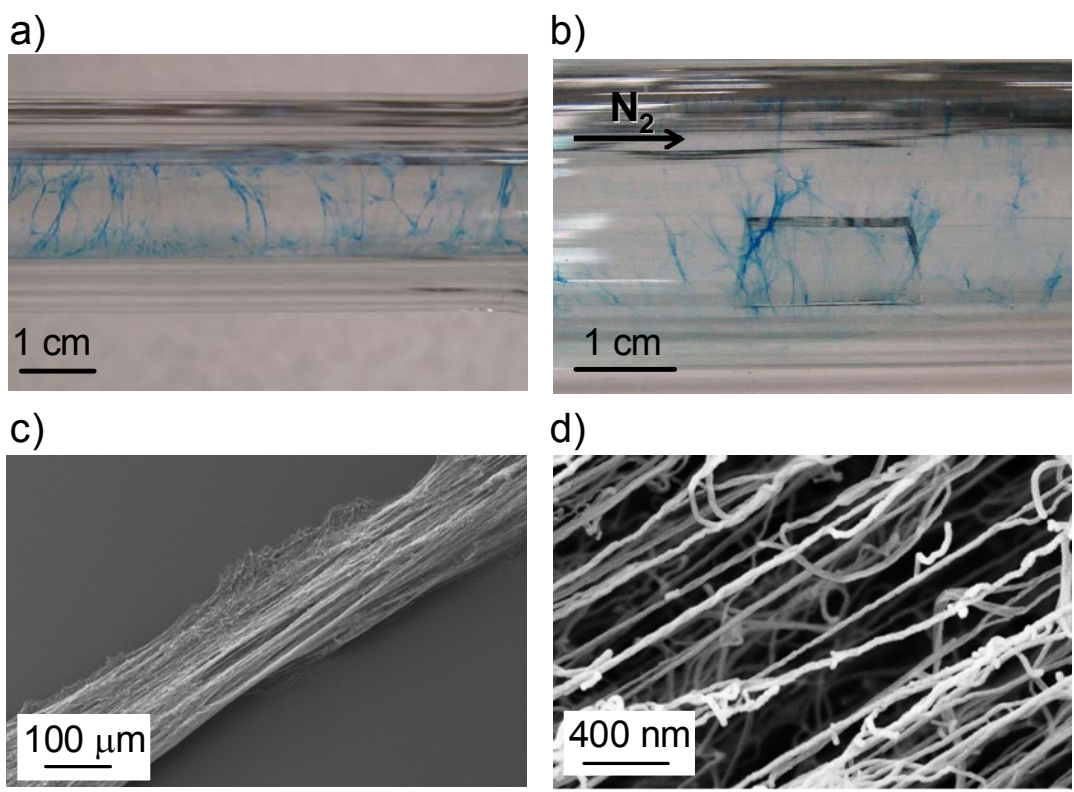

**Figure 1**

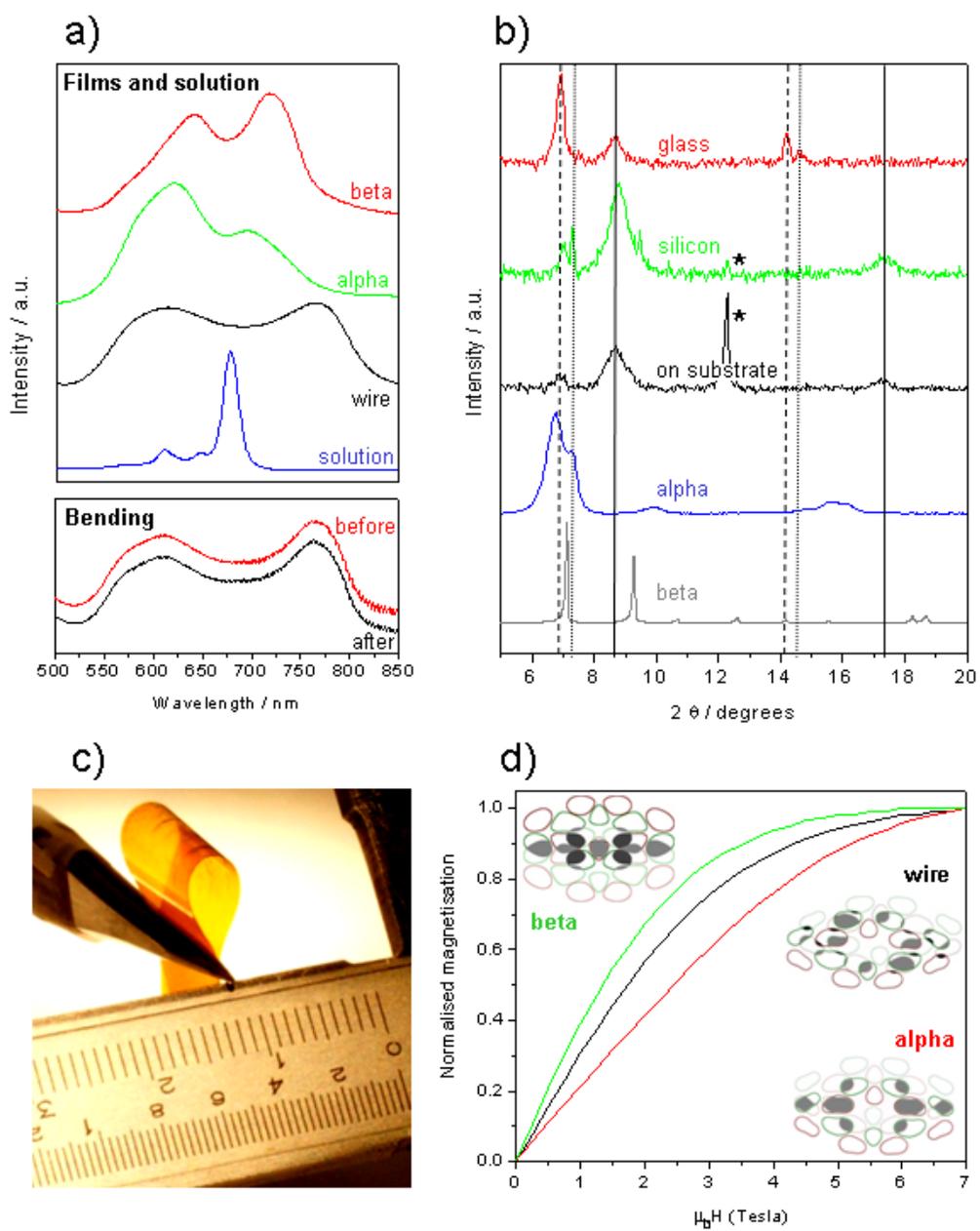

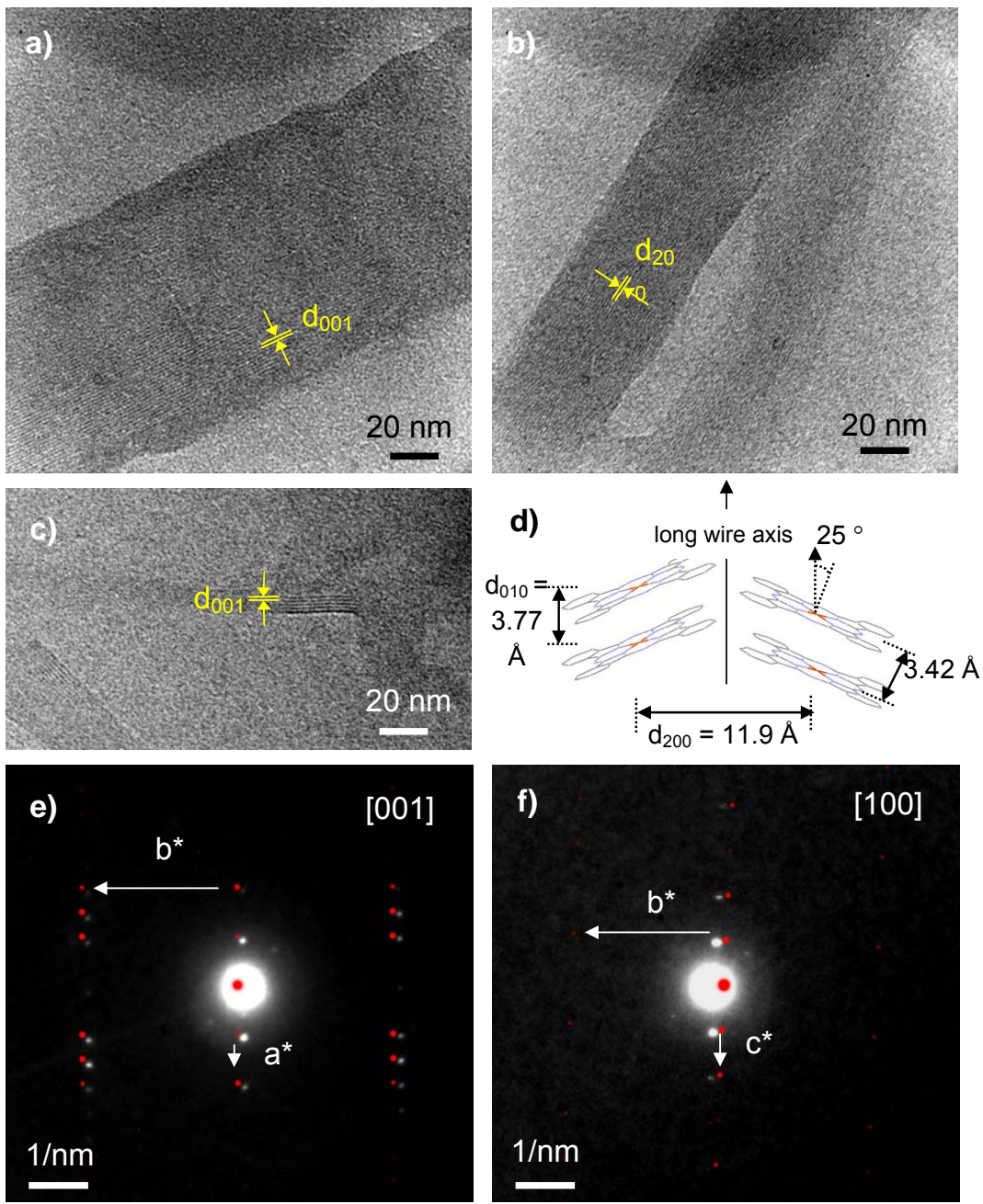

**Figure 3**

**Indexation of planes**

| h | k | l | XRD (Å) | TED (Å) | Calculated (Å) |
|---|---|---|---|---|---|
| 0 | 0 | 1 | 12.8 ± 0.2 | 12.6 ± 0.2 | 12.69 |
| 2 | 0 | 0 | 11.9 ± 0.2 | 12.0 ± 0.1 | 11.92 |
| 2 | 0 | -1 | 10.2 ± 0.1 |  | 10.20 |
| 0 | 1 | 0 |  | 3.77 ± 0.02 | 3.77 |

**Unit cell parameters**

|  | Nanowire | Alpha[11] | Beta[24] |
|---|---|---|---|
| Space Group | $P2_1/c$ | $P\bar{1}$ | $P2_1/a$ |
| Volume (Å$^3$) | 1187 | 582.3 | 1166 |
| a (Å) | 24.8 ± 0.4 | 12.886 | 19.407 |
| b (Å) | 3.77 ± 0.02 | 3.769 | 4.79 |
| c (Å) | 13.2 ± 0.2 | 12.061 | 14.628 |
| α (°) | 90 | 96.22 | 90 |
| β (°) | 106 ± 1 | 90.62 | 120.93 |
| γ (°) | 90 | 90.32 | 90 |
| Z | 2 | 1 | 2 |

**Table 1**

**Supporting Information**

The lattice potential energy (PE) was estimated by summing the bond potential $\phi_{ij}$ between every atom i in a reference molecule and every atom j in a neighbour, for the molecules whose contribution to the lattice PE was more than 1% of the total. $\phi_{ij}$ was calculated using the Lennard-Jones 9-6 form of the van der Waals interaction following the work by Yim et al. on Pc crystals:[1]

$$Lennard-Jones_{9-6}\phi_{ij} = \varepsilon_{ij}\left[2\left(\frac{R^*_{ij}}{R_{ij}}\right)^9 - 3\left(\frac{R^*_{ij}}{R_{ij}}\right)^6\right]$$

Since first row transition metal Pcs as well as metal-free phthalocyanine ($H_2Pc$) are isomorphous,[2] the central metal ion must have very little effect on the molecular packing and was disregarded in the calculation. Atoms in the Pc ring have very similar electronegativities so we assume coulombic contributions are negligible.

$R_{ij}^*$ is the minimum-energy separation between the $i^{th}$ and $j^{th}$ atoms respectively, and $\varepsilon_{ij}$ is the energy attained at $R_{ij}^*$. Values for homonuclear $R^*$ and $\varepsilon$ were obtained from molecular mechanic force field calculations[3] and are listed in Table 1.

| Atom | R* (Å) | ε (eV) |
|---|---|---|
| C | 4.00 | 0.00347 |
| N | 3.6 | 0.00694 |
| H | 2.8 | 0.00174 |

**Table 1** Lennard-Jones 9-6 parameters for homoatomic pairs [3].

For heteroatomic pairs we estimate $R^*$ to be the arithmetic mean of the respective homonuclear $R^*$'s and $\varepsilon$ to be the geometric mean of the respective homonuclear $\varepsilon$'s. $R_{ij}$ is the distance between non-bonded atoms and is the sum of the intramolecular distances and the separation between the molecules. Intramolecular distances were obtained from the single crystal XRD determined structure of β-CuPc.[4] The separation between molecules was derived from the lattice parameters that were calculated from XRD and electron diffraction data. The lattice PE was minimised using Solver in Excel by rotating the rigid molecules through ($\theta_x$, $\theta_y$, $\theta_z$) in a Cartesian framework as shown in Figure 1.

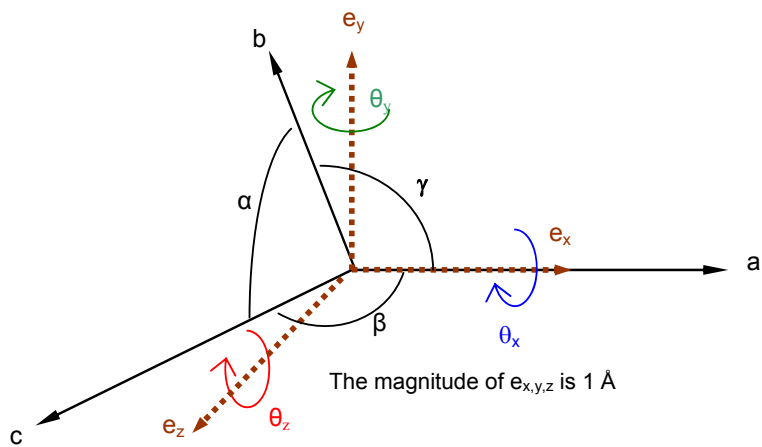

**Figure 1** Basis set used for rotating the molecule during energy minimisation.

The validity of our simple model was verified through the minimisation of β-CuPc which yielded a lattice PE minimised structure that was only ($\theta_x$ = 3.2°, $\theta_y$ = 2.4°, $\theta_z$ = -0.6°) away from the experimentally determined structure. The fractional coordinates of the atoms are listed in Table 2.

|     | a      | b      | c      |     | a      | b      | c      |
| --- | ------ | ------ | ------ | --- | ------ | ------ | ------ |
| Cu  | 0.000  | 0.000  | 0.00   | N1  | -0.128 | 0.358  | -0.076 |
| C1  | -0.098 | 0.332  | -0.146 | N2  | -0.046 | 0.195  | -0.130 |
| C2  | -0.121 | 0.445  | -0.255 | N3  | 0.014  | 0.120  | -0.246 |
| C3  | -0.172 | 0.594  | -0.309 | N4  | 0.058  | -0.090 | -0.070 |
| C4  | -0.181 | 0.671  | -0.414 | H1  | 0.105  | 0.017  | -0.353 |
| C5  | -0.139 | 0.597  | -0.466 | H2  | 0.198  | -0.223 | -0.339 |
| C6  | -0.089 | 0.452  | -0.412 | H3  | 0.258  | -0.458 | -0.170 |
| C7  | -0.080 | 0.372  | -0.305 | H4  | 0.220  | -0.507 | -0.014 |
| C8  | -0.033 | 0.225  | -0.225 | H5  | -0.204 | 0.649  | -0.269 |
| C9  | 0.056  | -0.021 | -0.172 | H6  | -0.220 | 0.788  | -0.458 |
| C10 | 0.108  | -0.126 | -0.193 | H7  | -0.147 | 0.658  | -0.548 |
| C11 | 0.129  | -0.101 | -0.280 | H8  | -0.057 | 0.399  | -0.452 |
| C12 | 0.182  | -0.229 | -0.271 |     |        |        |        |
| C13 | 0.216  | -0.369 | -0.175 |     |        |        |        |
| C14 | 0.195  | -0.395 | -0.087 |     |        |        |        |
| C15 | 0.142  | -0.268 | -0.098 |     |        |        |        |
| C16 | 0.108  | -0.244 | -0.023 |     |        |        |        |

**Table 2** Lattice energy minimised fractional atomic coordinates of η-CuPc